\documentclass[aip,apl,amsmath,reprint,superscriptaddress]{revtex4-1}%

\usepackage{graphicx}

\begin{document}

\title{Spectroscopy of mechanical dissipation in micro-mechanical membranes}

\author{Andreas J\"{o}ckel}
\affiliation{Departement Physik, Universit{\"a}t Basel, CH-4056 Basel, Switzerland}

\author{Matthew T. Rakher}
\affiliation{Departement Physik, Universit{\"a}t Basel, CH-4056 Basel, Switzerland}

\author{Maria Korppi}
\affiliation{Departement Physik, Universit{\"a}t Basel, CH-4056 Basel, Switzerland}

\author{Stephan Camerer}
\affiliation{Max-Planck-Institut f{\"u}r Quantenoptik and Fakult{\"a}t f{\"u}r Physik, Ludwig-Maximilians-Universit{\"a}t, 80799~M{\"u}nchen, Germany}

\author{David Hunger}
\affiliation{Max-Planck-Institut f{\"u}r Quantenoptik and Fakult{\"a}t f{\"u}r Physik, Ludwig-Maximilians-Universit{\"a}t, 80799~M{\"u}nchen, Germany}

\author{Matthias Mader}
\affiliation{Max-Planck-Institut f{\"u}r Quantenoptik and Fakult{\"a}t f{\"u}r Physik, Ludwig-Maximilians-Universit{\"a}t, 80799~M{\"u}nchen, Germany}

\author{Philipp Treutlein}\email[Electronic address: ]{philipp.treutlein@unibas.ch}
\affiliation{Departement Physik, Universit{\"a}t Basel, CH-4056 Basel, Switzerland}

\date{\today}

\begin{abstract}
We measure the frequency dependence of the mechanical quality factor ($Q$) of SiN membrane oscillators and observe a resonant variation of $Q$ by more than two orders of magnitude. The frequency of the fundamental mechanical mode is tuned reversibly by up to 40\% through local heating with a laser. Several distinct resonances in $Q$ are observed that can be explained by coupling to membrane frame modes. %, consistent with recent theoretical work. 
Away from the resonances, the background $Q$ is independent of frequency and temperature in the measured range.
%We can distinguish between clamping losses and other loss mechanisms, which are independent of frequency and temperature in the measurement range.
%We report on a method of tuning and analysis of the mechanical quality factor ($Q$) of micromechanical membrane oscillators. The membranes are tuned over up to 40\% of their frequency via local heating using a laser, causing the $Q$ factor to vary over two orders of magnitude. This change is analyzed and shown to be caused by coupling to mechanical modes of the frame. Using this technique, one can distinguish between these clamping losses and other mechanisms, which are independent of frequency and temperature in the measured intervals.
\end{abstract}

%\keywords{}

\maketitle
Micro-mechanical membrane oscillators are currently investigated in many optomechanics experiments, where lasers and optical cavities are used for cooling, control, and readout of their mechanical vibrations.\cite{Wilson09,Thompson08,Zwickl08,Camerer11,Friedrich11} Applications lie in the area of precision force sensing and in fundamental experiments on quantum physics at macroscopic scales.\cite{Kippenberg08} The quality factor $Q$ of the mechanical modes of the membranes is a key figure of merit in such experiments. However, the origin of mechanical dissipation limiting the attainable $Q$ is not completely understood and a subject of intense research.\cite{Verbridge07,Wilson-Rae08,Southworth09,Unterreithmeier10,Rae11,Cole11} 

Here we report an experiment in which we observe a variation of $Q$ by more than two orders of magnitude as a function of the fundamental mode frequency of a SiN membrane. Several distinct resonances in $Q$ are observed that can be explained by coupling to mechanical modes of the membrane frame.\cite{Wilson-Rae08,Rae11}  
%, as suggested by recent theoretical work.\cite{Wilson-Rae08}  
The frequency of the membrane modes is tuned reversibly by up to 40\% through local heating of the membrane with a laser. This method of frequency tuning has the advantage that the frequency dependence of $Q$ can be studied with a single membrane \textit{in situ}, resulting in a detailed spectrum of the coupling to the environment of this particular mode. Other methods that compare $Q$ between various structures of different sizes have to rely on the assumption that the environment of these structures is comparable.\cite{Cole11}

\begin{figure}
\centering
\includegraphics[width=1\columnwidth]{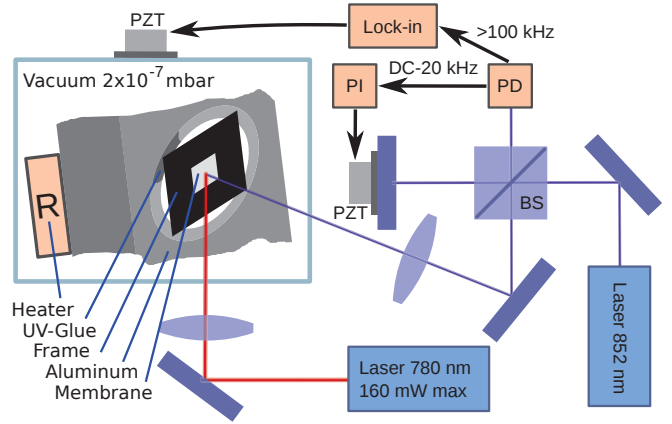}
\caption{\label{fig:paper1} Experimental setup. The SiN membrane in a Si frame is glued at one edge to an aluminum holder inside a room-temperature vacuum chamber. The heating laser (red) at 780~nm is power stabilized to $2 \times 10^{-4}$ RMS in a bandwidth of $12$~kHz, and focused onto the membrane under an angle.
%An imaging system on the back can be used to align the beam pointing onto the membrane. 
The membrane vibrations are read out with a stabilized Michelson interferometer (blue). The interferometer signal is also used for feedback driving of the membrane with a piezo (PZT). %The membrane amplitude is measured with a lock-in amplifier and driven with its integrated phase locked loop.
}
\end{figure}
%% Setup: Parts
We investigate ``low-stress'' SiN  membranes that are supported by a Si frame.\cite{cite:membrane} The frame is glued at one edge to a holder inside a vacuum chamber, see Fig.~\ref{fig:paper1}. The eigenfrequencies of a square membrane under tension are
\begin{eqnarray*}
f_{m,n} & = & \frac{1}{2 l}\sqrt{\frac{S}{\rho}\left(m^2+n^2 \right)},
\end{eqnarray*}
where $l$ is the side length, $\rho=2.9~\textrm{g}/\textrm{cm}^3$ the density,\cite{Zink04} and $S$ the tensile stress in the membrane. The modes are labeled by the number of anti-nodes $m$ and $n$ along the two dimensions. The stress $S=E \left(l-l_0 \right)/l_0$, where $E$ is  Young's modulus, arises in the fabrication process. The SiN membrane is stretched from its equilibrium length $l_0$ to the length $l$ of the Si frame.  
% of a displacement from its equilibrium length $l_0$, with the Young's modulus $E$, due to the fabrication process.

To read out the membrane vibrations, a Michelson interferometer operating at $852$~nm is used, where one end mirror consists of the membrane. The interferometer is stabilized by the DC to $20$~kHz part of the photodiode (PD) signal. The incident power on the membrane is $580~\mu$W in a diameter of $150~\mu$m and the position sensitivity is $1\times 10^{-14}~\textrm{m}/\sqrt{\textrm{Hz}}$. The $>100$~kHz frequency components of the signal %including the membrane signal 
are fed into a lock-in amplifier with integrated phase locked loop, which measures the membrane amplitude and drives its motion via a piezo mounted outside of the vacuum chamber.

%% Setup: Tuning
To tune the membrane frequency, a power stabilized $780$~nm laser is focused onto the membrane to a diameter of $350~\mu$m. This laser heats the membrane locally in its center. A second method of heating the whole membrane and frame is by a resistive heater (R) in the chamber. 

\begin{figure}[t]
\centering
\includegraphics[width=1\columnwidth]{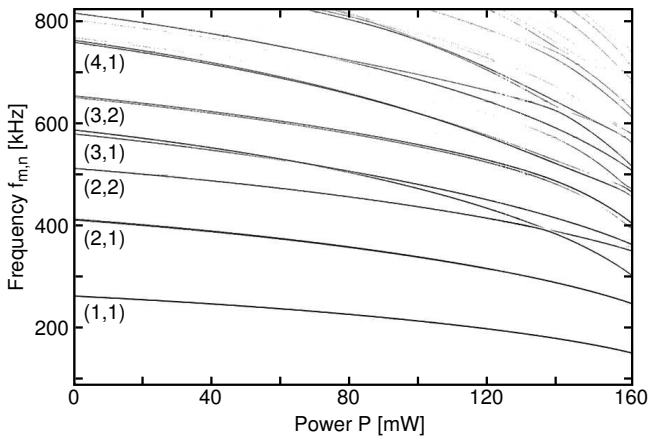}
\caption{\label{fig:paper2} Mode spectrum $f_{m,n}$ of a membrane ($l=0.5$~mm, thickness $t=50$~nm) as a function of $P$. At $P=0$, the lowest 13 modes lie within $2\%$ of the expected frequency. At higher $P$, anticrossings between higher order modes are visible.}
\end{figure}

%% Experiment: Frequency tuning
In a first experiment, we demonstrate the tunability of the membrane eigenfrequencies through laser heating. Fig.~\ref{fig:paper2} shows the recorded mode spectrum as a function of heating laser power $P$. The spectra are recorded by Fourier transforming the PD signal. One can see a reversible decrease of all mode frequencies $f_{m,n}$ with $P$.

The decrease in frequency can be attributed to a thermal expansion of the membrane $ \Delta l_0 /l_0=\alpha_{0}\Delta T+\alpha_{1} \Delta T^2$, where $\Delta l_0$ is the equilibrium length change and $\alpha_0$ ($\alpha_1$) the first (second) order expansion coefficient for a temperature change $\Delta T$. This reduces the  tensile stress by $\Delta S= -E \left(\Delta l_0 / l_0 \right)$. %with the Young's modulus $E$. 
In a simple model assuming a spatially homogeneous and linear temperature change with power $\Delta T= \chi P$, one can describe the power-dependence of the stress as
\begin{eqnarray*}
S & = &S_0 -E\left(\Delta l_0 / l_0 \right) =S_0-E\left(  \alpha_0 \chi P + \alpha_1 \chi^2 P^2 \right). 
\end{eqnarray*}
%A fit to the data for $f_{1,1}(P)$ 
A fit of $f_{1,1}(P)=\sqrt{a + bP + cP^2}$ to the data 
describes the observed dependence within $\pm 1$~kHz. For low $P$ we observe a linear shift of $\Delta f_{1,1}=-363~\mathrm{Hz}/\mathrm{mW}$. We neglect the dependence of $E$ on $\Delta T$ because it is small.\cite{Chuang04,Rouxel02} 
As shown in Tab.~\ref{tab:paper1}, the tunability of $f_{1,1}$ depends strongly on the geometry. 
% and $S_0$, which varies presumably due to small variations in the fabrication process. 

In order to extract $\chi$ from the fit, one has to measure $\alpha_0$. This is done by heating up the whole sample holder with the resistive heater. % as shown in Fig.~\ref{fig:paper1}. 
In this case both $l$ and $l_0$ change and the difference in the expansion coefficients $\Delta \alpha=\alpha_f-\alpha$ of the frame and the membrane determines $S-S_0=E\Delta\alpha \Delta T $. Heating the setup by $\Delta T=16$~K and using\cite{Chuang04} $E=260$~GPa and\cite{Lyon77} $\alpha_f=2.6$~ppm/K, one gets $\alpha_0=1.6$~ppm/K, $\alpha_1=1.3\times 10^{-8}/\textrm{K}^2$ and $\chi=0.6~\mathrm{K}/\mathrm{mW}$. This yields an average membrane temperature of $T=100^\circ$C for $P=160$~mW. 

To model laser absorption in the membrane, we perform a finite element (FEM) simulation of laser heating\cite{Wallquist10} using a Gaussian beam profile and a heat conductivity\cite{Zink04} $\kappa=3$~W/K\,m. From the resulting temperature distribution we calculate the average membrane temperature for a given absorbed laser power. By comparing with $\chi$, we find that a fraction of $1.5\times10^{-3}$ of the $780$~nm laser power is absorbed, an order of magnitude larger than the absorption in low-stress membranes at $1064$~nm.\cite{Thompson08,Zwickl08}

\begin{table}[t]
\begin{center}
\begin{tabular}{|l||r|r|r|r|r|r|}
\hline
$l$ [$\mu$m] & 250 & 500 & 1000 & 1500 & 500 & 500 \\ \hline
$t$ [nm] & 50 & 50 & 50 & 50 & 75 & 100 \\ \hline
$S_0$ [MPa] & 66.4 & 98.0 & 120 & 78.8 & 114 & 217 \\ \hline
$f_{1,1}$ [kHz] & 428 & 260 & 144 & 77.7 & 281 & 387 \\ \hline
$\Delta f_{1,1}$ [Hz/mW] & -259 & -363 & -68.9 & -49.5 & -89.6 & -10.5 \\ \hline
$Q_\mathrm{max}$ [$10^5$] & $3.2$ & $10$ & $15$ & $5.7^\star$ & $10$ & $0.37^\star $ \\ \hline
\end{tabular}
\end{center}
\caption{Summary of measured SiN membrane parameters. %$\Delta f$ refers to the frequency tuning using the laser at low powers. 
$Q_\mathrm{max}$ refers to the maximum observed $Q$. 
Values marked by $^\star$ were limited by the available tuning range.}
\label{tab:paper1}
\end{table}

%
%To determine the laser absorption in the membrane, we perform a finite element simulation of laser heating and the resulting stress change, using a Gaussian  beam profile, membrane heat conductivity\cite{Zink04} $\kappa=3$~W/K\,m, and fixed frame temperature. We find a peak temperature of $300^\circ$C in the membrane center for $P=160$~mW and an absorption of the laser light of $\approx 1.5\times10^{-3}$ at 780~nm. This is an order of magnitude larger than the absorption in other experiments employing longer wavelengths.\cite{Thompson08,Zwickl08}
%%At these high laser powers, we also observe annealing effects, which increase the membrane frequency permanently.
%%and heat capacity $C=400$~J/Kg/K

%Alpha_0m comes from frequency shift by frame heating and data  
%and extract the expansion coefficients... $\alpha_{m0}=1.9 10^{-6}/K\text{, }\alpha_{m1}=16.8 10^{-9}/K^2$. 

%\begin{eqnarray*}
%S-S_0 & = & \Delta T \left(\alpha_f-\alpha_m \right) E
%\end{eqnarray*}
%\begin{eqnarray*}
%f_0 & = & \frac{1}{l_m}\sqrt{\frac{S_0}{2\rho}}
%\end{eqnarray*}

\begin{figure}
\centering
\includegraphics[width=1\columnwidth]{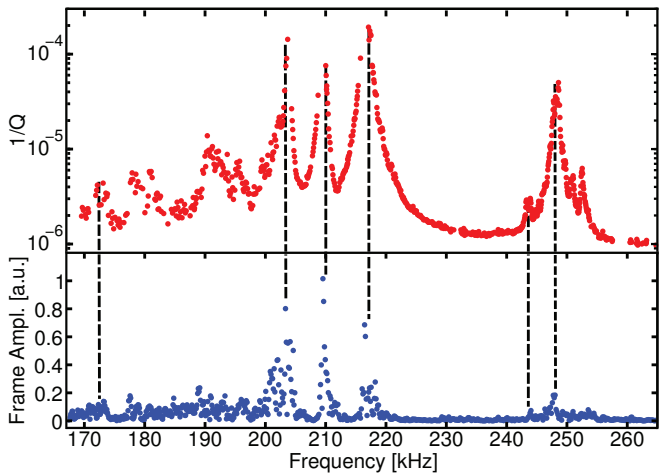}
\caption{\label{fig:paper3} Upper plot: spectrum of membrane dissipation $Q^{-1}(f_{1,1})$, showing a variation over two orders of magnitude. Lower plot: vibrations of the frame measured close to the membrane. The resonances in $Q^{-1}(f_{1,1})$ coincide with the frame modes. %, indicating coupling of the membrane vibrations to a (dissipative) frame mode.
}
\end{figure}

\begin{figure}
\centering
\includegraphics[width=1\columnwidth]{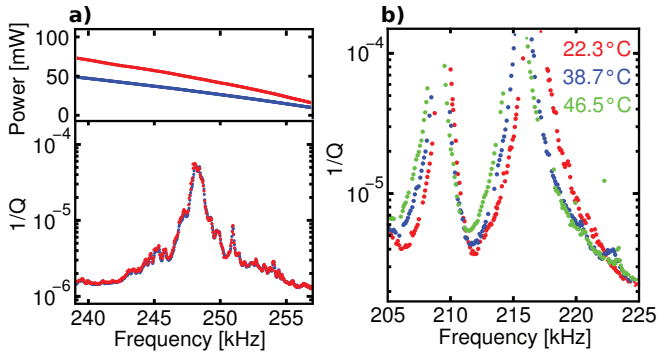}
\caption{\label{fig:paper3a} a) $f_{1,1}(P)$ and $Q^{-1}(f_{1,1})$ for different heating laser positions (membrane center: blue, off-center: red). The spectra $Q^{-1}(f_{1,1})$  overlap, indicating that $Q$ is directly dependent on frequency. 
b) $Q^{-1}(f_{1,1})$ for different sample holder temperatures. Heating shifts the frame modes to lower frequencies.}
\end{figure}

In a second experiment, we use laser tuning to record a spectrum of the quality factor $Q$ of the fundamental mode as a function of $f_{1,1}$.
We measure the decay time $\tau$ of the membrane amplitude in ring-down measurements after driving it to $\approx 0.5$~nm. The upper plot in Fig.~\ref{fig:paper3} shows the dissipation $Q^{-1}=1/\pi f \tau$. We observe distinct resonances, changing  $Q$ by more than two orders of magnitude. 
To show that the spectrum directly depends on $f_{1,1}$, %and not on $P$, 
the heating laser is pointed off center such that a different dependence $f_{1,1}\left(P\right)$ results, see Fig.~\ref{fig:paper3a}a. 
%The $Q$ factor still has the same dependence on $f_{1,1}$, showing that it only indirectly depends on $P$. % the membrane temperature.
The dependence $Q^{-1}(f_{1,1})$ is unchanged, showing that $Q$ only indirectly depends on $P$.
The resonances in $Q$ can be attributed to coupling of the membrane mode to modes of the frame. To prove this, the interferometer is pointed onto the frame next to the membrane and the amplitude response to a driving with the piezo is recorded, as shown in the lower plot in Fig.~\ref{fig:paper3}. The observed frame modes clearly overlap with the resonances in $Q^{-1}$. 
%In contrast to the independence of the membrane temperature on the spectrum, the frame has an influence. So far, the frame temperature was assumed to be constant, as it is much thicker than the membrane. 
If the frame is heated with the resistive heater, we observe a shift in the resonances in $Q^{-1}(f_{1,1})$, as shown in Fig.~\ref{fig:paper3a}b. We attribute this to a shift of the frame modes due to thermal expansion and decreasing Young's modulus.

All these measurements  prove that the coupling to frame modes  is responsible for the observed behavior of $Q$. A FEM simulation of the frame modes shows roughly the right density of modes in the frequency range of interest. As the eigenfrequencies depend strongly on the exact mounting, dimensions, and Young's modulus of the frame, it is difficult to model them quantitatively.

For stoichiometric $\textrm{Si}_3\textrm{N}_4$ ``high-stress'' membranes ($S_0=980$~MPa) we observe a much weaker dependence of the mode frequencies on $P$. The measurements indicate that absorption of $780$~nm light is lower by two orders of magnitude compared to the ``low-stress'' membranes. This is of importance for experiments coupling such membranes to atomic systems.\cite{Camerer11}  Using the limited tuning range of the resistive heater, we  also observe a change of $Q$ with frequency in high-stress membranes. This shows that coupling to frame modes is also important in this case. The highest measured $Q$ is $4\times 10^6$ for a high-stress membrane with $l=1.5$~mm and $t=50$~nm. 
%, as the heat transport increases only by a factor of ten \textbf{cite}.  %the heat transport 
%the heat transport is also enhanced 3 vs 20? w/m/k

%We also tried to structure low stress SiN membranes in order to get a lower mass using Focussed ion beam by cutting out a 50µm slab. However, these membranes $Q$ factor is limited to $10^5$ and their frequency is much more sensitive to the heating laser, limiting the maximum incident power to less than 10mW.
\begin{figure}
\centering
\includegraphics[width=1\columnwidth]{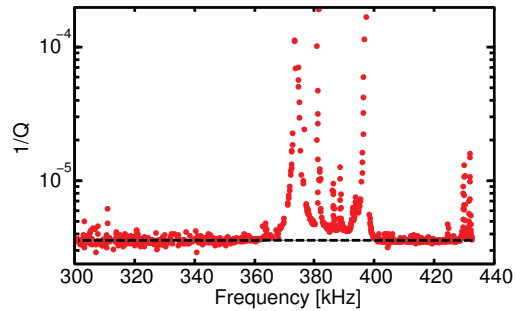}
\caption{\label{fig:paper4} Spectrum of membrane dissipation $Q^{-1}(f_{1,1})$ for another membrane ($l=250~\mu$m, $t=50$~nm). Besides  coupling to frame modes, the dissipation is independent of frequency.}
\end{figure}

Besides coupling to frame modes, the frequency dependence of other dissipation mechanisms is of interest. Fig.~\ref{fig:paper4} shows the dissipation spectrum of another low-stress membrane. Away from the resonances, we observe a constant baseline $Q_\mathrm{max}$, indicating that other dissipation mechanisms are independent of $f(S)$ and $T$ within our tuning range. This is in contrast to what has been observed in SiN strings.\cite{Verbridge07,Unterreithmeier10} We observe that $Q_\mathrm{max}$ increases with $l$, see Tab.~\ref{tab:paper1}. We also studied higher order modes up to $(2,2)$ and find approximately the same $Q_\mathrm{max}$, in contrast to other membrane experiments.\cite{Rae11} This could be due to the different frame geometry and mounting. In our case, the frame is a relatively small resonant structure with eigenmodes at distinct frequencies. 
%This shows that the clamping losses are only caused by coupling to localized frame modes, not propagating modes of the frame. 
This can be exploited to reduce clamping loss by tuning the membrane frequency to a gap between frame modes, analogous to the recently demonstrated phononic bandgap shielding.\cite{Alegre11}

In conclusion, we presented a precise method for laser-tuning of  micro-mechanical membrane oscillators and used it for spectroscopy of mechanical dissipation. Resonances in the dissipation were observed and explained as coupling to localized frame modes. Other dissipation mechanisms were found to be independent of  membrane frequency and temperature in the measured range. 

Our laser tuning technique could be extended to stoichiometric $\textrm{Si}_3\textrm{N}_4$ membranes by using a laser with smaller wavelength and thus higher absorption.\cite{Tan09} This would allow further investigation of the differences between low-stress and stoichiometric membranes. Moreover, it could be useful in finding optimal frame geometry and mounting conditions to circumvent clamping loss.
%This technique could also be used to probe the $Q$ factor at low temperatures and in general circumvent clamping losses in these setups.

We acknowledge helpful discussions with I. Wilson-Rae, M. Aspelmeyer, K. Hammerer, and T. W. H{\"a}nsch. Work supported by the EU project AQUTE and the NCCR Nanoscale Science.

%\bibliography{references.bib}

\begin{thebibliography}{10}%
\makeatletter
\providecommand \@ifxundefined [1]{%
 \ifx #1\undefined \expandafter \@firstoftwo
 \else \expandafter \@secondoftwo
\fi
}%
\providecommand \@ifnum [1]{%
 \ifnum #1\expandafter \@firstoftwo
 \else \expandafter \@secondoftwo
\fi
}%
\providecommand \enquote [1]{``#1''}%
\providecommand \bibnamefont  [1]{#1}%
\providecommand \bibfnamefont [1]{#1}%
\providecommand \citenamefont [1]{#1}%
\providecommand\href[0]{\@sanitize\@href}%
\providecommand\@href[1]{\endgroup\@@startlink{#1}\endgroup\@@href}%
\providecommand\@@href[1]{#1\@@endlink}%
\providecommand \@sanitize [0]{\begingroup\catcode`\&12\catcode`\#12\relax}%
\@ifxundefined \pdfoutput {\@firstoftwo}{%
 \@ifnum{\z@=\pdfoutput}{\@firstoftwo}{\@secondoftwo}%
}{%
 \providecommand\@@startlink[1]{\leavevmode}%
 \providecommand\@@endlink[0]{}%
}{%
 \providecommand\@@startlink[1]{%
  \leavevmode
  \pdfstartlink
   attr{/Border[0 0 1 ]/H/I/C[0 1 1]}%
   user{/Subtype/Link/A<</Type/Action/S/URI/URI(#1)>>}%
  \relax
 }%
 \providecommand\@@endlink[0]{\pdfendlink}%
}%
\providecommand \url  [0]{\begingroup\@sanitize \@url }%
\providecommand \@url [1]{\endgroup\@href {#1}{\urlprefix}}%
\providecommand \urlprefix [0]{URL }%
\providecommand \Eprint[0]{\href }%
\@ifxundefined \urlstyle {%
  \providecommand \doi [1]{doi:\discretionary{}{}{}#1}%
}{%
  \providecommand \doi [0]{doi:\discretionary{}{}{}\begingroup
  \urlstyle{rm}\Url }%
}%
\providecommand \doibase [0]{http://dx.doi.org/}%
\providecommand \Doi[1]{\href{\doibase#1}}%
\providecommand \bibAnnote [3]{%
  \BibitemShut{#1}%
  \begin{quotation}\noindent
    \textsc{Key:}\ #2\\\textsc{Annotation:}\ #3%
  \end{quotation}%
}%
\providecommand \bibAnnoteFile [2]{%
  \IfFileExists{#2}{\bibAnnote {#1} {#2} {\input{#2}}}{}%
}%
\providecommand \typeout [0]{\immediate \write \m@ne }%
\providecommand \selectlanguage [0]{\@gobble}%
\providecommand \bibinfo [0]{\@secondoftwo}%
\providecommand \bibfield [0]{\@secondoftwo}%
\providecommand \translation [1]{[#1]}%
\providecommand \BibitemOpen[0]{}%
\providecommand \bibitemStop [0]{}%
\providecommand \bibitemNoStop [0]{.\EOS\space}%
\providecommand \EOS [0]{\spacefactor3000\relax}%
\providecommand \BibitemShut [1]{\csname bibitem#1\endcsname}%
%</preamble>
\bibitem{Thompson08}%
  \BibitemOpen
  \bibfield{author}{%
  \bibinfo {author} {\bibfnamefont{J.~D.}\ \bibnamefont{{Thompson}}} \bibinfo
  {author} {\textit{et al.}},\ }%
  \bibfield{journal}{%
  \Doi{10.1038/nature06715}{\bibinfo {journal} {Nature}}\ }%
  \textbf{\bibinfo {volume} {452}},\ \bibinfo {pages} {72} (\bibinfo {year}
  {2008}).%
  %\bibAnnoteFile{NoStop}{Thompson08}%
\bibitem{Zwickl08}%
  \BibitemOpen
  \bibfield{author}{%
  \bibinfo {author} {\bibfnamefont{B.~M.}\ \bibnamefont{{Zwickl}}} \bibinfo
  {author} {\textit{et al.}},\ }%
  \bibfield{journal}{%
  \Doi{10.1063/1.2884191}{\bibinfo {journal} {Appl. Phys. Lett.}}\ }%
  \textbf{\bibinfo {volume} {92}},\ \bibinfo {pages} {103125} (\bibinfo {year}
  {2008}).%
  %\bibAnnoteFile{NoStop}{Zwickl08}%
\bibitem{Wilson09}%
  \BibitemOpen
  \bibfield{author}{%
  \bibinfo {author} {\bibfnamefont{D.~J.}\ \bibnamefont{{Wilson}}}, \bibinfo
  {author} {\bibfnamefont{C.~A.}\ \bibnamefont{{Regal}}}, \bibinfo {author}
  {\bibfnamefont{S.~B.}\ \bibnamefont{{Papp}}},\ and\ \bibinfo {author}
  {\bibfnamefont{H.~J.}\ \bibnamefont{{Kimble}}},\ }%
  \bibfield{journal}{%
  \Doi{10.1103/PhysRevLett.103.207204}{\bibinfo {journal} {Phys. Rev. Lett.}}\
  }%
  \textbf{\bibinfo {volume} {103}},\ \bibinfo {pages} {207204} (\bibinfo {year}
  {2009}).%
  %\bibAnnoteFile{NoStop}{Wilson09}%
\bibitem{Camerer11}%
  \BibitemOpen
  \bibfield{author}{%
  \bibinfo {author} {\bibfnamefont{S.}~\bibnamefont{{Camerer}}} \bibinfo
  {author} {\textit{et al.}},\ }%
  %\bibfield{journal}{%
  %\bibinfo {journal} {ArXiv e-prints}}%
  % (\bibinfo {year} {2011}),\
  \Eprint{http://arxiv.org/abs/1107.3650}{arXiv:1107.3650} (\bibinfo {year}{2011}).
  %\bibAnnoteFile{NoStop}{Camerer11}%
\bibitem{Friedrich11}%
  \BibitemOpen
  \bibfield{author}{%
  \bibinfo {author} {\bibfnamefont{D.}~\bibnamefont{{Friedrich}}} \bibinfo
  {author} {\textit{et al.}},\ }%
  %\bibfield{journal}{%
  %\bibinfo {journal} {ArXiv e-prints}}%
  % (\bibinfo {year} {2011}),\
  \Eprint{http://arxiv.org/abs/1104.3251}{arXiv:1104.3251} (\bibinfo {year}{2011}).
  %\bibAnnoteFile{NoStop}{Friedrich11}%
\bibitem{Kippenberg08}%
  \BibitemOpen
  \bibfield{author}{%
  \bibinfo {author} {\bibfnamefont{T.~J.}\ \bibnamefont{{Kippenberg}}}\ and\
  \bibinfo {author} {\bibfnamefont{K.~J.}\ \bibnamefont{{Vahala}}},\ }%
  \bibfield{journal}{%
  \Doi{10.1126/science.1156032}{\bibinfo {journal} {Science}}\ }%
  \textbf{\bibinfo {volume} {321}},\ \bibinfo {pages} {1172} (\bibinfo {year}
  {2008}), %
  \bibinfo{author}{\bibfnamefont{F.}~\bibnamefont{Marquardt}} \bibnamefont{and}
  \bibinfo{author}{\bibfnamefont{S.~M.} \bibnamefont{Girvin}},
  \bibinfo{journal}{Physics} \textbf{\bibinfo{volume}{2}}, \bibinfo{pages}{40}
  (\bibinfo{year}{2009}), 
\bibinfo{author}{\bibfnamefont{I.}~\bibnamefont{Favero}} \bibnamefont{and}
  \bibinfo{author}{\bibfnamefont{K.}~\bibnamefont{Karrai}},
  \bibinfo{journal}{Nat. Photonics} \textbf{\bibinfo{volume}{3}},
  \bibinfo{pages}{201} (\bibinfo{year}{2009}), 
  D. Hunger \textit{et al.}, arXiv:1103.1820 (2011)
  %\bibAnnoteFile{NoStop}{Kippenberg08}%
  
  \bibitem{Verbridge07}%
  \BibitemOpen
  \bibfield{author}{%
  \bibinfo {author} {\bibfnamefont{S.~S.}\ \bibnamefont{{Verbridge}}}, \bibinfo
  {author} {\bibfnamefont{D.~F.}\ \bibnamefont{{Shapiro}}}, \bibinfo {author}
  {\bibfnamefont{H.~G.}\ \bibnamefont{{Craighead}}},\ and\ \bibinfo {author}
  {\bibfnamefont{J.~M.}\ \bibnamefont{{Parpia}}},\ }%
  \bibfield{journal}{%
  \Doi{10.1021/nl070716t}{\bibinfo {journal} {Nano Letters}}\ }%
  \textbf{\bibinfo {volume} {7}},\ \bibinfo {pages} {1728} (\bibinfo {year}
  {2007}).%
  %\bibAnnoteFile{NoStop}{Verbridge07}%
  
  \bibitem{Wilson-Rae08}%
 I.~Wilson-Rae, Phys. Rev. B \textbf{77}, 245418 (2008).
   
\bibitem{Southworth09}%
 D. R. Southworth \textit{et al.}, Phys. Rev. Lett. \textbf{102}, 225503 (2009).
   
\bibitem{Unterreithmeier10}%
 Q. P. Unterreithmeier, T. Faust, and J. P. Kotthaus, Phys. Rev. Lett. \textbf{105}, 027205 (2010).
 
  
  \bibitem{Rae11}%
  \BibitemOpen
  \bibfield{author}{%
  \bibinfo {author} {\bibfnamefont{I.}~\bibnamefont{{Wilson-Rae}}} \bibinfo
  {author} {\textit{et al.}},\ }%
  \bibfield{journal}{%
  \Doi{10.1103/PhysRevLett.106.047205}{\bibinfo {journal} {Phys. Rev. Lett.}}\
  }%
  \textbf{\bibinfo {volume} {106}},\ \bibinfo {pages} {047205} (\bibinfo {year}
  {2011}).%,\ \Eprint{http://arxiv.org/abs/1010.2171}{arXiv:1010.2171
  %[cond-mat.mes-hall]}%
  %\bibAnnoteFile{NoStop}{Rae11}%
  
  
\bibitem{Cole11}%
  \BibitemOpen
  \bibfield{author}{%
  \bibinfo {author} {\bibfnamefont{G.~D.}\ \bibnamefont{{Cole}}} \bibinfo
  {author} {\textit{et al.}},\ }%
  \bibfield{journal}{%
  \Doi{10.1038/ncomms1212}{\bibinfo {journal} {Nature Communications}}\ }%
  \textbf{\bibinfo {volume} {2}} \bibinfo {pages} {231} (\bibinfo {year} {2011}).\ %
  %\bibAnnoteFile{NoStop}{Cole11}%
\bibitem{cite:membrane}%
  \BibitemOpen
  \bibinfo {title} {www.norcada.com}\ %
  %\bibAnnoteFile{NoStop}{cite:membrane}%
\bibitem{Zink04}%
  \BibitemOpen
  \bibfield{author}{%
  \bibinfo {author} {\bibfnamefont{B.}~\bibnamefont{{Zink}}}\ and\ \bibinfo
  {author} {\bibfnamefont{F.}~\bibnamefont{{Hellman}}},\ }%
  \bibfield{journal}{%
  \Doi{10.1016/j.ssc.2003.08.048}{\bibinfo {journal} {Solid State
  Commun.}}\ }%
  \textbf{\bibinfo {volume} {129}},\ \bibinfo {pages} {199} (\bibinfo {year}
  {2004}).%
  %\bibAnnoteFile{NoStop}{Zink04}%
\bibitem{Chuang04}%
  \BibitemOpen
  \bibfield{author}{%
  \bibinfo {author} {\bibfnamefont{W.-H.}\ \bibnamefont{Chuang}}, \bibinfo
  {author} {\bibfnamefont{T.}~\bibnamefont{Luger}}, \bibinfo {author}
  {\bibfnamefont{R.}~\bibnamefont{Fettig}},\ and\ \bibinfo {author}
  {\bibfnamefont{R.}~\bibnamefont{Ghodssi}},\ }%
  \bibfield{journal}{%
  \Doi{10.1109/JMEMS.2004.836815}{\bibinfo {journal} {J. Microelectromech. S.
}}\ }%
  \textbf{\bibinfo {volume} {13}},\ \bibinfo {pages} {870 } (\bibinfo {year}
  {2004}).%,\ ISSN \bibinfo {issn} {1057-7157}%
  %\bibAnnoteFile{NoStop}{Chuang04}%
\bibitem{Rouxel02}%
  \BibitemOpen
  \bibfield{author}{%
  \bibinfo {author} {\bibfnamefont{T.}~\bibnamefont{Rouxel}} \bibinfo {author}
 {\textit{et al.}},\ }%
  \bibfield{journal}{%
  \Doi{10.1016/S1359-6454(02)00004-6}{\bibinfo {journal} {Acta
  Mater.}}\ }%
  \textbf{\bibinfo {volume} {50}},\ \bibinfo {pages} {1669} (\bibinfo {year}
  {2002}).%,\ ISSN \bibinfo {issn} {1359-6454}%
  %\bibAnnoteFile{NoStop}{Rouxel02}%
    
\bibitem{Lyon77}%
 K.~G.~Lyon \textit{et al.}, J. Appl. Phys. \textbf{48}, 865 (1977).
 
\bibitem{Wallquist10}%
M.~Wallquist \textit{et al.}, Phys. Rev. A \textbf{81}, 023816 (2010).

 

\bibitem{Alegre11}%
  \BibitemOpen
  \bibfield{author}{%
  \bibinfo {author} {\bibfnamefont{T.~P.~M.}\ \bibnamefont{Alegre}}, \bibinfo
  {author} {\bibfnamefont{A.}~\bibnamefont{Safavi-Naeini}}, \bibinfo {author}
  {\bibfnamefont{M.}~\bibnamefont{Winger}},\ and\ \bibinfo {author}
  {\bibfnamefont{O.}~\bibnamefont{Painter}},\ }%
  \bibfield{journal}{%
  \Doi{10.1364/OE.19.005658}{\bibinfo {journal} {Opt. Express}}\ }%
  \textbf{\bibinfo {volume} {19}},\ \bibinfo {pages} {5658} (\bibinfo {year}
  {2011}).%
  %\bibAnnoteFile{NoStop}{Alegre11}%
\bibitem{Tan09}%
  \BibitemOpen
  \bibfield{author}{%
  \bibinfo {author} {\bibfnamefont{W. C.}~\bibnamefont{Tan}}, \bibinfo
  {author} {\bibfnamefont{S.}~\bibnamefont{Kobayashi}}, \bibinfo
  {author} {\bibfnamefont{T.}~\bibnamefont{Aoki}}, \bibinfo
  {author} {\bibfnamefont{R. E.}~\bibnamefont{Johanson}}, and\ \bibinfo
  {author} {\bibfnamefont{S. O.}~\bibnamefont{Kasap}},\ }%
  \bibfield{journal}{%
 {\bibinfo {journal} {J Mater. Sci.: Mater. Electron.}}\ }%
  \textbf{\bibinfo {volume} {20}},\ \bibinfo {pages} {S15} (\bibinfo {year}
  {2009}).%,\ ISSN \bibinfo {issn} {0927-0248}%
  %\bibAnnoteFile{NoStop}{Tan09}%
\end{thebibliography}

\bibliographystyle{apsrev4-1}

\end{document}